# A Pathway between Bernal and Rhombohedral Stacked Graphene Layers with Scanning Tunneling Microscopy

P. Xu,[1] Yurong Yang,[1,2] D. Qi,[1] S. D. Barber,[1] M. L. Ackerman,[1] J. K. Schoelz,[1] T. B. Bothwell,[1] Salvador Barraza-Lopez,[1] L. Bellaiche,[1] and P. M. Thibado[1,a]

[1]Department of Physics, University of Arkansas, Fayetteville, Arkansas 72701, USA

[2]Physics Department, Nanjing University of Aeronautics and Astronautics, Nanjing 210016, China

Horizontal shifts in the top layer of highly oriented pyrolytic graphite, induced by a scanning tunneling microscope (STM) tip, are presented. Excellent agreement is found between STM images and those simulated using density functional theory. First-principle calculations identify that the low-energy barrier direction of the top layer displacement is toward a structure where none of the carbon $p_z$ orbitals overlap, while the high-energy barrier direction is toward AA stacking. Each directional shift yields a real-space surface charge density similar to graphene; however the low-energy barrier direction requires only one bond length to convert ABA (Bernal) to ABC (rhombohedral).

---

[a] Electronic mail: thibado@uark.edu

The unusual band structure of graphene is one of its most fascinating features.[1] The linear dispersion of the *E-k* relation near the six vertices of the Brillouin zone results in charge carriers that behave like massless Dirac fermions.[2,3] The giant electron charge density and high-carrier mobility at room temperature make graphene an excellent candidate for high-frequency transistors[4] inside flexible electronics. One of the chief obstacles, however, for using graphene in electronic devices is its lack of a band gap at the *K*-point. Recent work has pointed toward stacking graphene layers as a way to solve this problem. Trilayer graphene is especially interesting because two stable allotropes have been identified; the layers can be arranged with ABA (Bernal) stacking or ABC (rhombohedral) stacking. This is an exciting new research area primarily due to the significant differences in their electronic properties resulting from such a subtle alteration.[5-8] Specifically, ABC trilayers exhibit an inherent band gap of ~6 meV at the *K*-point,[9] which can be increased by applying an electric field,[10] while no such band gap is predicted in ABA trilayers.

Naturally, several major steps have been taken toward characterizing the stacking sequence. For instance, Raman spectroscopy performed on mechanically exfoliated graphene has revealed that the majority of the trilayers produced are ABA stacked, while only about 15% are in the ABC configuration.[11,12] On the other hand, when graphene is grown on SiC(0001), the layers selectively form in the ABC order over ABA, as observed with high-resolution transmission electron microscopy.[13] Certainly, one would like to control the stacking sequence or ideally alter it from one form to the other. A related area with a lot of interest is rotated or twisted layers. This has a lot of appeal, because all the physics can be parameterized with just one angle.[14] Horizontal shifting has received less attention. In this Letter, we present scanning tunneling microscopy (STM) images on highly oriented pyrolytic graphite (HOPG) which show clear evidence of the top layer shifting in a particular horizontal direction. In addition, we find excellent agreement with a series of density functional theory (DFT) simulated STM images generated from structures shifted along this same direction. From DFT we also find that the lowest-energy barrier for transitioning from ABA to ABC stacking is found to be in this same direction. Finally, within one STM image we have caught a transition from a graphite-like surface charge density to a graphene-like surface charge density and this data shows a horizontal shift in the top layer has occurred in the same direction and by the same amount.

Experimental STM images were obtained using an Omicron ultrahigh-vacuum (base pressure is $10^{-10}$ mbar), low-temperature STM operated at room temperature. The top layers of a 6 mm × 12 mm × 2 mm thick piece of HOPG were exfoliated with tape to expose a fresh surface. It was then mounted with silver paint onto a flat tantalum sample plate and transferred through a load-lock into the STM chamber, where it was electrically grounded. STM tips were electrochemically etched from 0.25 mm diameter tungsten wire via a custom double lamella setup with an automatic cutoff.[15] After etching, they were gently rinsed with distilled water, briefly dipped in a concentrated hydrofluoric acid solution to remove surface oxides, and then loaded into the STM chamber. All STM images (filled-state) were collected at a constant current of 0.1 nA and a tip bias of +0.1 V.

Nine panels are shown in Fig. 1, with the top row showing three characteristic atomic-resolution STM images, the middle row showing three DFT simulated STM images, and the bottom row showing ball-and-stick models with different amounts of top-layer horizontal shifting. The first STM image highlights the equilateral triangular lattice typical for ABA-stacked HOPG in Fig. 1(a). Here only every other carbon atom in the top layer is being imaged. A similar image is expected for ABC-stacked graphite, but since ABA-stacked is the most common naturally occurring form, we consider this our starting configuration.[16] A second STM image illustrating a vertical row-like structure is shown in Fig. 1(b). A honeycomb structure characteristic of graphene can be seen in Fig. 1(c). Anyone that has done STM on HOPG has observed at one time or another STM images that are not the typical outcome shown in Fig. 1(a). We have systematically cataloged all of these images into groups in order to see a pattern. For example, in certain areas of the surface we can scan the tip along the carbon-carbon bond axis direction and the image changes from the triangular pattern to the row-like pattern shown in Fig. 1(b). To ensure this was not due to tip asymmetry, but was due to the sample, this same tip was repositioned to other areas of the sample (a few mm away) and the experiment was repeated. We found the effect was associated with that specific region of the sample and did not follow the tip. Another event being reported in the literature is the observation of a graphene-like surface charge density on graphite as shown in Fig. 1(c).[17]

To gain insight into the origin of these results, simulated STM images of graphite were extracted from DFT calculations without modeling the STM tip.[18,19] These calculations were performed within the local-density approximation to DFT using projector-augmented wave potentials[20] as implemented in the plane-wave basis set VASP[21] code. The graphite was modeled as a six-layer Bernal stack and a cutoff energy of 500 eV. A very large $160 \times 160 \times 1$ Monkhorst-Park $k$-point mesh was used to ensure proper sampling around the Dirac point. All forces were less than 0.1 eV/nm, resulting in a carbon-carbon bond length of 0.142 nm and an interplanar separation of 0.334 nm. With the top layer of the graphite in its equilibrium position a simulated STM image was generated as shown in Fig. 1(d) and a ball-and-stick model for the top two layers is shown in Fig. 1(g). To best replicate the experimental STM conditions, the local density of states was integrated from the Dirac point to 0.2 eV below that point, and an appropriate isocontour surface was chosen. Excellent agreement is obtained with the STM data shown in Fig. 1(a). The simulated image shows a significant decrease in the charge density for half the carbon atoms. This is known to be due to half of the top layer carbon atoms forming a "dimmer bond" with the carbon atom directly underneath it in the plane below.[22] This "bonding" is responsible for reducing the surface charge density of these atoms which is then detected with the STM.

Next, in the computer model, the top layer of the graphite was shifted 0.30 bond lengths along the carbon-carbon (C-C) bond-axis and a new STM image was simulated as shown in Fig. 1(e). A row-like structure similar to Fig. 1(b) can be seen. The stacking arrangement for this configuration is best observed in the model shown in Fig. 1(h). One can see how the overlap between the carbon atoms in the different layers is responsible for this row-like structure. For a shift of 0.50 bond lengths, the charge density around all sites has equalized, as seen in Fig. 1(f). This gives the nice honeycomb structure and has excellent agreement with the STM data shown in Fig. 1(c). For the ball-and-stick model shown in Fig. 1(i), the top-layer carbon atoms are now centered over the benzene ring in the layer below. We define this symmetric configuration, as the "no overlap" structure, since none of the carbon $p_z$ orbitals overlap with the ones in the plane below. The "no overlap" structure was pointed out before.[23] Note, the AA stacking also yields a graphene-like surface charge density,[16] but we argue that the horizontal shift is not in this direction.

Further evidence that the pathway between ABA and ABC is happening along the bond-axis direction and toward the "no overlap" structure is presented in Fig. 2. The energy per carbon atom found from DFT while shifting the top layer in ten equally spaced steps from ABA to ABC is shown in Fig. 2(a). Note, at each step the top-layer was allowed to relax perpendicular to the surface to find the lowest energy pathway. The top layer is shifted along the C-C bond axis direction but directly toward the "no overlap" configuration. For this shift the ABC arrangement occurs after one bond length, and the "no overlap" situation is at the halfway point. Notice the symmetric shape of the energy curve indicates that there is no difference between the top surface layer shifting from ABA toward "no overlap" and from ABC toward "no overlap." When the top layer is shifted in the opposite direction the DFT energy per atom is shown in Fig. 2(b). When shifting in this direction the top layer must move 2 bond lengths before reaching the ABC stacking and the halfway point is the well-known AA stacking configuration. Notice, the barrier height for the "no overlap" direction is about 1.5 meV/atom, while the opposite direction yields a barrier height of about 15 meV/atom. It should be noted that the van der Waals interaction was not included in these calculations. However, we also repeated the calculations including the van der Waals interaction following the prescription of Román-Pérez and Soler.[24] The outcome is similar to those shown in Fig. 2(a) and (b), except the energy barrier height was slightly reduced (~15%). Ball-and-stick models illustrating the stacking arrangement for the ABA, halfway point, and ABC are also shown. Halfway through the shift shown in Fig. 2(a) is the point where the no overlap structure occurs between the top two layers [as more clearly illustrated in Fig. 1(i)] and a graphene-like surface charge density exists. Notice that the energy curve has a relatively flat top at this point, which may result in a meta-stable state, allowing the STM tip to occasionally image this higher-energy configuration. In fact, we believe the electrostatic attraction to the STM tip helps stabilize this configuration.[25,26]

An STM image of the surface is shown as an inset in Fig. 2(a). This image has "caught" the sudden transition from a graphite-like surface charge density to a graphene-like surface charge density. On the left side of the image is the signature triangular lattice of graphite, while on the right side is the distinct honeycomb structure of graphene. As mentioned earlier, previous work has demonstrated graphene formation on HOPG;[23] however, it was manifested as a continuous transition across the image. Our image is the result of a sudden jump, with the fast scan direction oriented vertically, in which a discrete change has occurred about one-third of the way through (the image was flattened and cropped for clarity). Note the previous scan in this location showed all graphite while the subsequent scan showed all graphene. Overlaid on top of the image are two ball-and-stick models of the graphene structure. The left-hand model is fit to the graphite image, while the right-hand model is fit to the graphene image. The overlapping region of the two models shows the best-fit horizontal displacement that has occurred during the movement of the top layer. This shows that the top layer of the graphite was, in fact, shifted in the direction previously discussed, and by an amount that corresponds to half a bond length.

By analyzing the coordination number of the top layer atoms with the second layer atoms, one can understand why the "no overlap" configuration should have the lowest energy graphene-like surface charge density. In both the ABA and ABC stacks, half the atoms have 6 nearest neighbors (with the plane below) and half the atoms have only 1, for an average coordination number of 3.5. When in the "no overlap" configuration the atoms each have 3 nearest neighbors, for an average coordination of 3. When in the "AA stacking" configuration the atoms each have 1 nearest neighbor, for an average coordination of 1.

In conclusion, we have presented experimental evidence for a horizontal shift in the top layer of a graphite sample, induced by an STM tip and resulting in the observation of a graphene-like surface charge density. This shift is direction dependent, a conclusion supported by first principles DFT energy calculations and simulated STM images. Specifically, the lowest-energy barrier direction is toward "no overlap" of the $p_z$ orbitals, while the highest-energy barrier direction is toward "AA stacking" (full overlap of the $p_z$ orbitals). The observed shift from a graphite-like surface charge density to a graphene-like surface charge density is a signature that the top layer has shifted halfway along a path between ABA and ABC stacking.

P.X. and P.T. gratefully acknowledge the financial support of the Office of Naval Research (ONR) under grant number N00014-10-1-0181 and the National Science Foundation (NSF) under grant number DMR-0855358. Y.Y. and L.B. thank ARO Grant W911NF-12-1-0085, the Office of Basic Energy Sciences, under contract ER-46612 for personnel support and NSF grants DMR-0701558 and DMR-1066158; and ONR Grants: N00014-11-1-0384 and N00014-08-1-0915 for discussions with scientists sponsored by these grants. Calculations were made possible thanks to the MRI grant 0959124 from NSF, N00014-07-1-0825 (DURIP) from ONR, and a Challenge grant from HPCMO of the U.S. Department of Defense.

**FIGURE CAPTIONS**

FIG. 1 (Color online). (a-c) Filled-state STM images of the HOPG surface collected at a constant current of 0.1 nA and a tip bias of +0.1 V. (d-f) Simulated STM images of graphite extracted from DFT with the top layer shifted horizontally by a fraction of the C–C bond length indicated. (g-i) Ball-and-stick structural models showing the locations of the carbon atoms in the top two layers of graphite for a horizontal shift of 0.00, 0.30, and 0.50 bond lengths, respectively.

FIG. 2 (Color online). (a) DFT energy/atom is plotted (squares) as a function of the top-layer horizontal shift (away from AA stacking), in the direction illustrated by the three ball-and-stick models shown. The pink frame represents the top layer, the green frame the second layer, and the blue frame is the third layer. Inset image: STM data showing graphite- and graphene-like surface

charge densities in a single scan. Ball-and stick models are overlaid on the graphite and graphene surfaces to indicate the locations of the carbon atoms as well as the direction and magnitude of the horizontal shift that occurred (the white box behind the models highlights the shift between them). (b) DFT energy/atom is plotted (circles) as a function of the top-layer horizontal shift (toward AA stacking), which is in the direction indicated by the three ball-and-stick models shown. At each step, the atoms were allowed to relax perpendicular to the plane before determining the associated energy and simulating the STM image.

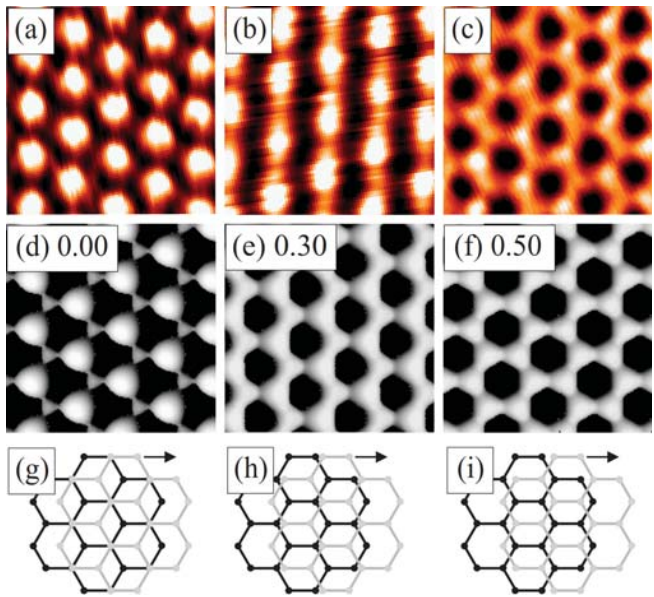

Fig.1    by XU et al.

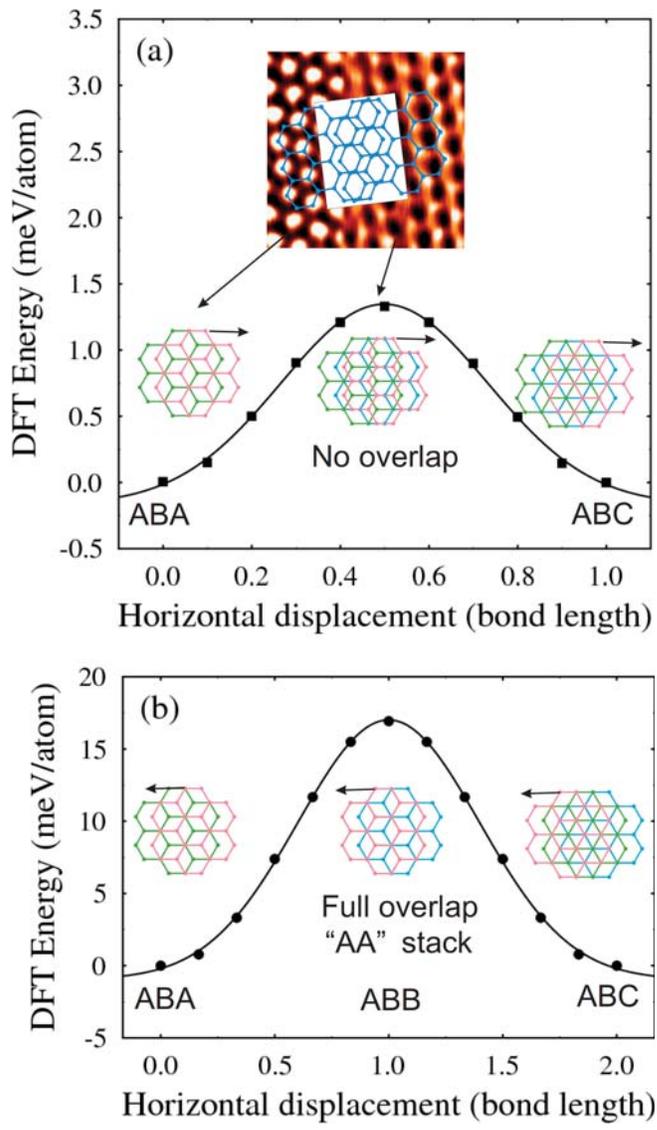

Fig.2    by XU et al.